\documentclass[lettersize,journal]{IEEEtran}

\usepackage{amsmath,amsfonts}
\usepackage{algorithmic}
\usepackage{algorithm}
\usepackage{array}
\usepackage[caption=false,font=normalsize,labelfont=sf,textfont=sf]{subfig}
\usepackage{textcomp}
\usepackage{stfloats}
\usepackage{url}
\usepackage{verbatim}
\usepackage{graphicx}
\usepackage{cite}
\usepackage{xcolor}
\usepackage{hyperref}
\usepackage{lipsum}
\usepackage{multicol}
\usepackage{xcolor}
\usepackage{soul}
\usepackage{caption}
\usepackage{tabularx,booktabs,rotating}
\usepackage{placeins}
\usepackage{multirow}
\usepackage{caption}    
\usepackage{needspace}  
\usepackage{cite}
\usepackage{ragged2e}
\usepackage{multirow, tabularx, graphicx}
\usepackage{flushend, multicol}
\begin{document}
\title{The 5P Reflection Model for Education in the Generative Artificial Intelligence (GenAI) Era}
\author{Rajan Kadel$^1$, Samar Shailendra$^2$, Islam Mohammad Tahidul$^2$, Urvashi Rahul Saxena$^2$, Aakanksha Sharma$^2$, and Sabitra Kaphle$^3$
~\IEEEmembership{\\$^1$School of IT, National Academy of Professional Studies (NAPS), Australia}
~\IEEEmembership{\\$^2$School of IT and Engineering, Melbourne Institute of Technology (MIT), Australia}
~\IEEEmembership{\\$^3$School of Health, Medical and Applied Sciences, CQUniversity, Australia}}       
\markboth{This is the pre-print version of the paper submitted to IEEE Transactions on Education}%
{Kadel \MakeLowercase{\textit{et al.}}: The 5P Reflection Model for Education in the Generative Artificial Intelligence (GenAI) Era}

\maketitle

\begin{abstract}
Contributions: A reflection model suitable for the era of Generative  Artificial Intelligence (GenAI) is introduced. The proposed model is an integrated model that extracts features from various existing models and also incorporates technological aspects of GenAI. 

Background: Universities worldwide are facing challenges in adopting GenAI into their curricula, as it has impacted academic integrity and the scholarship of teaching and research. Traditional reflection models are struggling to authenticate student reflection as GenAI is incorporated in education. This requires a GenAI-aware model to enable the opportunities that address the associated challenges with GenAI.

Research Questions: Does the academic ecosystem require a GenAI-aware reflection model to adopt GenAI into education? How to make a reflection model structured to ensure student authenticity and cognitive engagement within a GenAI-aware learning environment?

Methodology: This study employs a design-based research methodology, assisted by a critical inquiry approach, to analyse existing reflection models in the era of GenAI and examine the technological aspects of GenAI. Additionally, it identifies a gap and the absence of a comprehensive reflection model that enhances reflection in the era of GenAI and supports the effective integration of GenAI in education.

Findings: The literature survey indicates there is a greater need for reflection in the era of GenAI to ensure learning, but the existing reflection models lack proper strategies to manage the problem introduced by GenAI. A standard GenAI-aware reflection model, called \textit{5P (Purpose, Process, Product, Pitfalls and Plan)}, is proposed i) to manage greater demand for reflection, ii) to address the limitations of the existing reflection models, iii) to consider technological aspects of GenAI. 
\end{abstract}

\section{Introduction} \label{sec:introduction}
The rapid adoption of Generative Artificial Intelligence (GenAI) in higher education has created new opportunities for learning, creativity, and productivity. However, it also raises concerns around plagiarism, over-reliance, misinformation, and ethical boundaries of use in learning and research. The key research questions guiding this study have emerged with the need to reassess reflective practice models within the education system, particularly as learners increasingly rely on GenAI tools for interpretation, writing, and understanding.

To remain relevant in the GenAI era, reflection models must address several key requirements such as focus on process than the product, transparency of the reflective process, learners interactions with AI tools, the prompts being used, and unwind the iterative thinking that unfolds through human‑AI collaboration. It must also incorporate mechanisms to evaluate authenticity by distinguishing learners’ own reasoning from AI‑generated contributions, reinforce ethical and responsible AI usage, such as principles of attribution, integrity, and disclosure, must be embedded within the model rather than treated as external considerations. Additionally, a contemporary reflection model must account for the technological mediation of thinking, recognising how algorithms, interfaces, and equitable access of AI shape the reflective experience. Without these elements, reflection frameworks will continue to misalign with the emerging behaviours of learners who increasingly co‑create meaning with GenAI. To address these emerging challenges, this study intends to answer the following research questions that examine the limitations of existing models and explore how a new, contemporary model can better support authentic, in-depth, and ethically sound reflection in GenAI-rich learning environments.
\begin{itemize}
\item What are the limitations of applying traditional reflective practice models in GenAI-rich learning environments?
\item How can a new reflection model be designed to enhance authenticity and in-depth learning in the era of GenAI?
\end{itemize}

In response to these questions, this paper proposes a reflection model that re-centres the student within the technological loop. The development of this model follows a Design-Based Research (DBR) methodology~\cite{dbrc2003design} assisted by Critical Inquiry~\cite{brookfield1995becoming}, ensuring that the model is both theoretically grounded and practically applicable. A detailed discussion of this two-stage development process covering problem identification and theoretical synthesis is provided in Section~\ref{sec:primer}. Through advocating ``process over product'' philosophy, this approach aims to foster meaningful reflection for personal and professional growth while navigating the complexities of GenAI-integrated learning.

This research culminates in a new reflection model titled \textbf{\textit{5P (Purpose, Process, Product, Pitfalls and Plan)}} reflection model. The \textit{5P} model synthesises essential features from established pedagogical frameworks while reviewing specific mechanisms existing in the literature to promote the effective and responsible use of GenAI. The model comprehensively considers GenAI technology's strengths and weaknesses and provides a structured approach for authentic, transparent, and ethically sound reflection.

The outline of the paper is as follows: Section~\ref{sec:TF} provides an overview of the current reflection practices and associated challenges and limitations in the era of GenAI. Section~\ref{sec:RW} provides a brief review of the recent work on the reflection model and its importance in the era of GenAI. The basis and steps for developing a new model are presented in Section~\ref{sec:primer}. The proposed 5P reflection model and discussion with adoption considerations are presented in Section~\ref{sec:P5} and Section~\ref{sec:Discussion}, respectively. Finally, Section~\ref{sec:conclusions} concludes the paper.

\section{Theoretical Foundation} \label{sec:TF}
Reflection is widely acknowledged as a cornerstone of critical pedagogy and effective learning. Various types of reflection models have been created to organise and improve the reflective process. Reflective practices span multiple theoretical traditions. In this study, we classify reflection into five commonly recognised categories~\cite{moon2013handbook,finlay2008reflecting,mann2009reflection}: Experiential reflection (Kolb \cite{kolb1984experiential}; Gibbs \cite{gibbs1988learning}; Boud et al. \cite{boudreflection,boud2013reflection}), transformative reflection (\cite{mezirow1991transformative,Mezirow2009}; Brookfield et al. \cite{brookfield1995becoming, brookfield2017becoming}), emotional (affective) reflection (Moon \cite{moon2013handbook}), reflective practitioner approaches (Schön \cite{schon1983reflective,Schon1987}), and self-regulated reflective processes aligned with metacognitive and self regulated learning frameworks (Zimmerman \cite{zimmerman2000attaining}; Pintrich \cite{pintrich2000role}; Winne \cite{winne2000measuring}). In this section, we will provide a brief overview of various types of reflection models in these learning practices and perform an analysis of using these models in the era of GenAI. 

\subsection{Experiential Learning Based Models}\label{sec:experiential_models}

\begin{table*}[htbp]
\caption{A summary analysis of experiential learning based models.}
\centering
\setlength{\tabcolsep}{6pt}
\renewcommand{\arraystretch}{1.2}
\begin{tabular}{p{1.3cm} p{5.6cm} p{5.4cm} p{3.5cm}}
\hline
\textbf{Model} & \textbf{Strengths} & \textbf{Limitations - GenAI} & \textbf{Application Area(s)} \\\hline
Kolb (1984) \cite{kolb1984experiential} &  Connects theory and practice; supports iterative experiential learning & GenAI may interrupt learner ownership of reflection; risks abstraction detached from lived experience & STEM, management, and design education \\ \hline

Gibbs (1988) \cite{gibbs1988learning} & Encourages deep critical reflection; integrates emotions; supports iterative improvement & GenAI may oversimplify emotional reflection; can lead to formulaic responses when AI-generated prompts dominate & Nursing, education, business, and health sciences \\ \hline

DIEP (1985) \cite{boudreflection} &  Promotes metacognitive awareness; supports structured reflective writing and forward planning & GenAI may dominate the planning stage; limited emotional depth in AI-generated outputs & Health sciences, teaching, and social sciences \\ \hline
\end{tabular}
\label{tab:reflection_models_experirential}
\end{table*}

Experiential learning reflection models emphasise learning through real-world experience followed by structured reflection to connect action with deeper understanding and improvement. One of the early models in this category is Kolb’s experiential learning model, which conceptualises learning as a cyclical process in which experience is systematically transformed into knowledge through reflection, conceptual understanding, and action \cite{kolb1984experiential}. Through interaction with experience, the model emphasises learning as an adaptable and iterative process that allows learners to continuously improve their comprehension. 
The process of experiential learning begins with a concrete experience, where a learner’s direct participation in an activity serves as the primary foundation for knowledge. This is followed by reflective observation, a stage where the individual deliberately reviews the experience from multiple perspectives to process actions and outcomes. These reflections are then transformed during abstract conceptualisation, where the learner connects their insights to existing frameworks or professional standards to understand the underlying principles and patterns of why events occurred. Finally, through active experimentation, the learner applies these new concepts to real-world situations, testing strategies and learning from both successes and failures to refine their skills and restart the cycle of continuous improvement.

Another popular experiential framework is Gibbs’ Reflective Cycle, a practical model designed to help learners deeply analyse their personal experiences \cite{gibbs1988learning}. Unlike models that focus solely on actions, Gibbs’ cycle captures a holistic view by examining how the learner felt, why the event unfolded as it did, and what specific steps can be taken if a similar situation arises in the future. Gibbs’ Reflective Cycle begins with Description, where the learner provides a clear, non-judgemental account of what happened and who was involved. This is followed by Feelings, a phase dedicated to self-awareness where the learner explores their emotions both during the event and while reflecting. In the Evaluation phase, the learner judges the experience to identify what went well and what did not, which then feeds into the Analysis phase. Here, the focus shifts from what happened to why it happened, using professional theories and evidence to investigate the underlying causes of the situation. The cycle concludes by translating these insights into growth through the Conclusion and Action Plan stages. During the conclusion, the learner summarises their takeaways, identifying personal skills that need enhancement and determining what could have been done differently. Finally, the action plan ensures that the reflection leads to iterative improvement by defining specific strategies and approaches to be implemented if a similar event occurs in the future. 

Another experiential model popular in educational settings is the Describe, Interpret, Evaluate, and Plan (DIEP) reflection model \cite{boudreflection}. The DIEP model begins with Describe, where the learner provides an objective, quantitative account of an incident to separate factual data from emotional reactions. This leads into the Interpret stage, where the practitioner critically analyses these facts to gain new insights and experiences, a process frequently utilised in behavioural and sentiment analysis. To address the often complex nature of measuring learning, the Evaluate stage performs a baseline review of the activity to quantify the specific benefits or usefulness acquired from the experience. Finally, the Plan stage converts these evaluative insights into specific, measurable, and time-bound professional development strategies, ensuring that the reflection results in enhancements for future performance.

Table~\ref{tab:reflection_models_experirential} provides an overview of experiential learning based reflection models, their strengths, limitations in the era of GenAI and their primary applications. Interaction with GenAI tools (e.g., prompting, iteration, and refinement) can be explicitly treated as an active and concrete learning experience requiring systematic reflection. Key features adopted from these experiential models include staged reflection, cyclical learning through action and feedback, and forward-oriented improvement. However, these models are unable to handle the various challenges introduced by the GenAI technology.  

\subsection{Transformative Learning Based Models}\label{sec:TL}

\begin{table*}[htbp]
\caption{A summary analysis of transformative based models.}
\centering
\setlength{\tabcolsep}{6pt}
\renewcommand{\arraystretch}{1.2}
\begin{tabular}{p{1.5cm} p{5.5cm} p{5.8cm} p{3.8cm}} \hline
\textbf{Model} &  \textbf{Strengths} & \textbf{Limitations} & \textbf{Application Area(s)} \\ \hline
Mezirow (1991) \cite{mezirow1991transformative} & Established the core framework of the theory (10 phases of transformation), emphasizing ``disorienting dilemmas'' as the catalyst for change & Heavily biased toward cognitive/rational processing, largely neglecting the role of emotion, intuition, and spirituality in the original formulation & Adult education, personal development, counselling/ psychology \\ \hline

Mezirow and Taylor (2009)\cite{Mezirow2009} & Incorporates diverse perspectives—emotional, spiritual, neurobiological, and cultural—addressing earlier critiques of rationality & The shift to diverse contexts makes it harder to standardise ``what counts'' as transformation; risks diluting the core concept  &  Community development, workplace learning, higher education, social justice education \\ \hline

McClain (2024) \cite{McClain2024} & Specifically addresses contemporary issues like digital learning environments, inclusivity, and the evolving definition of ``transformation'' in a polarised world & The attempt to unify such a fragmented field may struggle to reconcile the deep contradictions between different approaches (e.g., Critical Theory vs. Neurobiology) & Contemporary adult education, digital/online learning environments, inclusive and diverse educational settings \\ \hline
\end{tabular}
\label{tab:reflection_models_TL}
\end{table*}

Transformative learning, originally conceptualised by Jack Mezirow, positions reflection as a deep, critical, and often disruptive process through which learners re-examine their assumptions, beliefs, and habitual ways of understanding the world \cite{mezirow1991transformative}. Unlike surface-level or descriptive reflection, transformative learning emphasises perspective transformation, a fundamental shift in how individuals make meaning from their experiences. In educational and professional contexts, such transformation is particularly valuable because it supports the development of adaptive expertise, ethical awareness, critical thinking, and professional identity. 

Transformative learning in practice is proposed by Mezirow et al. \cite{Mezirow2009}, which explicates the core stages of individual perspective transformation as a process initiated by a disorienting dilemma. This critical event forces a re-evaluation of one's deeply held assumptions, known as meaning perspectives. The individual stages involve (i) critical self-reflection, (ii) rational discourse with others to test the validity of assumptions, and ultimately, (iii) a changed point of view or habit of mind. A similar but updated work in \cite{McClain2024} presents the foundational, cognitive-rational model of change. 
However, its primary limitation is its potential overemphasis on the rational-critical model, often marginalising the affective, emotional, and spiritual dimensions of learning. Furthermore, it has been criticised for being overly individualistic, thereby neglecting the collective and systemic contexts that profoundly shape and constrain the transformative process. Critics also note that the theory can appear overly deterministic and positive, failing to account for regressive or non-emancipatory outcomes of critical reflection. This aligns transformative learning closely with reflective practice traditions in teaching, nursing, counselling, engineering, management, and other professional fields where practitioners must revisit and reshape their thinking to address complex or uncertain situations.

A change from a rationalist foundation to a pluralistic, context-based application characterises the development of transformative learning, as given in Table \ref{tab:reflection_models_TL} for all three models. Although it was critiqued for ignoring emotion, Mezirow's 1991 model, which focused on the cognitive process of critical reflection prompted by a disorienting dilemma, laid the foundation. The model was enhanced in 2009 with the concept by incorporating environmental and emotional elements. Lastly, McClain's 2024 development aims to unify several contemporary viewpoints on transformation by attempting a meta-theoretical unification that addresses the complexity and fragmentation brought about by digital and social transformations. Despite this evolution, these models still do not adequately address the specific challenges introduced by GenAI.

\subsection{Reflective Practitioner Based Models}\label{sec:ReflectivePractioner}

\begin{table*}[htbp]
\caption{A summary analysis of reflective practitioner based models.}
\centering
\setlength{\tabcolsep}{6pt}
\renewcommand{\arraystretch}{1.2}
\begin{tabular}{p{1.5cm} p{5cm} p{5.5cm} p{4.5cm}}
\hline
\textbf{Model} & 
\textbf{Strengths} &
\textbf{Limitations} &
\textbf{Application Areas} \\
\hline
Schön (1983) \cite{schon1983reflective,Schon1987} & Focuses on reflection-in-action; emphasises professional competence and tacit knowledge; highly conceptual. &  Lacks thorough step-by-step instructions; inadequate emphasis on external social/cultural influences.& Training expert practitioners; professional development; transforming research into professional knowledge.\\ \hline

Brookfield (1995) \cite{brookfield1995becoming} & Employs numerous lenses to examine preconceived notions; features strongly on critical reflection and power dynamics.& Time-consuming; needs to seek critical feedback. & Higher education: exploring stereotypes and biases; improving democratic teaching methods.\\ \hline

Ghaye \& Lillyman (2010) \cite{ghaye2010reflection} & Emphasises reflection in supervision and action research, including sharing reflections with others&  Requires immense organisational support; can be quite challenging due to integration with action research and change management& Team development and organisational change; structured reflective supervision.\\ \hline

Rolfe et al. (2001) \cite{rolfe2001critical} & Step-by-step, cyclic, and easy to learn; establishes a clear progression from description to future action.&  Likely to promote superficial reflection if probing inquiries are not used; primarily focuses on reflection-on-action.& Entry-level training and education; incident reporting; and a brief, systematic review of simple events\\ \hline

 CARL (2011) \cite{edinburghCARL} & Sequential analysis separates context, action, and results clearly to determine Learning; action-outcome linkage is strong & Utilised for retrospective assessment rather than deep, ethical reflection; can be too mechanical if not applied critically & Preparing for structured performance reviews or interviews; linking specific actions to outcomes.\\ \hline
\end{tabular}
\label{tab:reflection_models_RP}
\end{table*}
Schön's model promotes professional development by modifying how we learn through our work \cite{schon1983reflective,Schon1987}. It focuses on three key components: knowledge-in-action, reflection-in-action, and reflection-on-action. then defining them in bullets or paragraph.
Industry professionals involved in constantly framing and reframing problems can potentially uncover inefficiencies and develop innovative context-specific solutions. This approach extends beyond technical ability, fostering self-awareness as well as emotional intelligence, which is imperative for leadership. The model serves as an essential tool for avoiding stagnation, ensuring that decision-making remains agile, relevant, and continuously improving \cite{schon1983reflective,Schon1987}. The three core components of Schön’s model are: 

\begin{itemize}
    \item Knowledge in Action: This component helps create a professional ``instinct'' through training and experience. It facilitates smooth decision-making for practitioners without having to actively pause and ponder each step.
    \item Reflection in Action: This component is popularly described as ``thinking on your feet''. It empowers an individual to critique and redefine behaviour during the occurrence of an event. It ensures flexibility and receptiveness to the changing circumstances.
    \item Reflection on Action: This component is also perceived as the ``post-game analysis'' that occurs after the event. It involves retrospection to evaluate what worked, gain new insights, and plan improvements for the future.\\
\end{itemize}

Brookfield and Ghaye \& Lillyman \cite{brookfield1995becoming,ghaye2010reflection} emphasise social reflection, thereby challenging practitioners to critically examine practice via external lenses and within organisational contexts. On the contrary, Context-Action-Results-Learning (CARL)~\cite{edinburghCARL}, Jasper's Experience-Reflection-Action (ERA)~\cite{jasper2013reflection}, and Rolf ~\cite{rolfe2001critical} models are cyclically designed and structured practical tools that facilitate step-by-step review of post-events. The CARL model thrives at integrating particular actions to quantitative outcomes for clear learning; however, ERA and Rolf models have gained popularity for their simplicity. Overall, a reflective practitioner selects a model based upon the its usage purpose, conceptual breadth \cite{schon1983reflective}, critical insight \cite{brookfield1995becoming}, and/or structured, actionable review \cite{edinburghCARL,jasper2013reflection}. Table~\ref{tab:reflection_models_RP} presents a summarised analysis of various reflective practitioner based models. Although these models incorporate effective reflection processes, they still lack the necessary capabilities to address the challenges posed by GenAI.

\subsection{Emotional (Affective) Reflection Based Models} \label{sec:emotional}
\begin{table*}[b]
\caption{A summary analysis of emotional (affective) based reflection models.}
\centering
\setlength{\tabcolsep}{6pt}
\renewcommand{\arraystretch}{1.2}
\begin{tabular}{p{1.5cm} p{5.5cm} p{5cm} p{4.5cm}}
\hline
\textbf{Model} & 
\textbf{Strengths} &
\textbf{Limitations} &
\textbf{Application Area(s)} \\
\hline

Boud, Keogh \& Walker's reflection \cite{boudreflection} & Provides a widely used structured process (returning, attending to feelings, re-evaluating) for post-experience learning & Focuses heavily on the individual process; may overlook social and institutional barriers to effective reflection & Adult and higher education, particularly for facilitating learning from real-world tasks. \\
\hline
Schutz \& DeCuir’s model \cite{schutz2002inquiry,schutz2002introduction} & Schutz and DeCuir focus on how students and teachers use personal goals and beliefs as reference points to evaluate their progress, triggering emotions within social-historical contexts. & Primarily targeted at emotion only, limiting its direct application to while using GenAI in education, & Researchers use this framework to understand emotions in various contexts, like teacher stress, student anxiety, and emotional regulation.  \\
\hline

Fook \& Gardner’s model \cite{Fook2007} & Emphasises critical analysis of power dynamics, assumptions, and social context within practice & Demands a high level of self-awareness and theoretical understanding, which can be challenging for standard users, & Social work, professional supervision, and challenging inequitable practice in organisations.  \\
\hline
\end{tabular}
\label{tab:reflection_models_emotion}
\end{table*}

Emotional reflection models recognise that emotions are integral to experience and learning, not separate from them. These models focus on how practitioners attend to, understand, and integrate feelings (both positive and negative) to deepen insight and guide ethical decision-making in practice \cite{boudreflection,bulman2013reflective,Fook2007}. D. Boud et al. \cite{boudreflection} outlined a model of reflection as a three-stage process: (a) returning to the experience, (b) attending to feelings, and (c) re-evaluating the experience. This establishes reflective practice as a deliberate, structured process for turning raw experience into transformative learning, emphasising the necessity of critically reviewing, validating emotions, and re-evaluating actions rather than simply describing events \cite{Schon1987}. It is predominantly used in health science professional development \cite{Mann2009} and professional management \cite{Cunliffe2016}, where managing ethical dilemmas and emotional responses (features central to Boud's emphasis on acknowledging feelings) is paramount. 

A key feature is the move beyond purely rational analysis; a disadvantage is the subjectivity inherent in deeply introspective processes. This reflective practice is characterised by its iterative, non-linear complexity, focusing heavily on the internal cognitive and affective (emotional) processes, making it highly accessible for self-directed professional learners but often considered less measurable and scalable for formal assessment compared to rigid models. These models tend to emphasise outcomes and challenges, while offering limited support for documenting the reflective process and personal growth. In the era of GenAI, the core challenge is that the automation of synthesis can bypass the emotional and critical ``stickiness'' required for genuine reflection, substituting GenAI-generated insights for personal cognitive struggle, leading to superficial or mimicked reflection \cite{Denton2020,Raza2023}. 

Table \ref{tab:reflection_models_emotion} illustrates the evolution of reflective practice where Boud et al. offered a structured process for learning from experience. Bulman and Schutz applied reflection specifically to clinical settings. Fook and Gardner advanced the field by emphasising critical reflection, focusing on power dynamics and social context within professional practice. Despite the emotional factor being crucial in GenAI-era learning, these models are not equipped with mechanisms to address GenAI’s technological aspects.

\subsection{Self-Regulated Learning Based Models} \label{sec:SRL}

\begin{table*}[htbp]
\caption{A summary analysis of self-regulated learning based models.}
\centering
\setlength{\tabcolsep}{6pt}
\renewcommand{\arraystretch}{1.2}
\begin{tabular}{p{2cm} p{5.5cm} p{5.5cm} p{3cm}}
\hline
\textbf{Model} & 
\textbf{Strengths} &
\textbf{Limitations} &
\textbf{Application Area(s)} \\
\hline

Zimmerman’s model (1990) \cite{zimmerman1990self} & Easy to integrate with GenAI tools for planning and checklist; guided self-reflection and adjustment in learning strategies & May skip genuine planning or reflection by letting GenAI generate steps; risk of over-reliance may reduce metacognitive effort & Health sciences, STEM education, and professional development contexts\\
\hline
Pintrich’s Model (2000) \cite{pintrich2000role} & GenAI can support multi-domain monitoring (motivation check-ins, cognitive strategies, contextual adaptation) & Higher complexity makes GenAI scaffolding hard to personalise; GenAI may hamper learning interpretation and judgement & Teacher education, research \\ \hline
Winne \& Hadwin Model (2000) \cite{winne2000measuring} & Perfect for GenAI-enhanced metacognitive prompts and real-time analytics; GenAI can track hints, planning, revision patterns to support learning & Possible to outsource tactical decisions to GenAI instead of learning strategy formation; Risks losing human judgement & Learning analytics research, STEM, problem-solving domains \& software development\\ \hline
Boekaerts’ dual-processing model (1999) \cite{boekaerts1999self}& Suitable for learners with emotional or motivational vulnerabilities; focus on well-being and emotion & More focus on well-being and emotion; less on problem-solving domains & Social-emotional learning, counselling programs, \& emotion-aware pedagogy \\
\hline
\end{tabular}
\label{tab:reflection_models_SRL}
\end{table*}
Self-regulated learning posits that effective learners are not passive recipients of knowledge, rather self-directed agents who regulate their cognition, motivation, and behaviour to optimise performance \cite{zimmerman1990self}. Self-regulated learning is a metacognitive process wherein learners actively manage their learning through goal setting, strategic planning, monitoring, and reflection. These models draw heavily on social-cognitive theory, highlighting the interplay between personal factors (beliefs, self-efficacy), behavioural strategies (planning, time management), and environmental influences (feedback, learning context). Zimmerman framed the cyclical self-regulated learning model (forethought, performance, self-reflection) \cite{zimmerman1990self}, whereas Pintrich’s (2000) self-regulated learning model \cite{pintrich2000role} is organised into four phases: forethought, monitoring, control, and reflection, and examines regulation across four domains: cognition, motivation/affect, behaviour, and context. Pintrich’s model is more comprehensive than Zimmerman's model, as it highlights learners as active agents who set goals, manage strategies, adapt to changing demands, and evaluate their progress. Thess models are widely used in educational research to understand how students develop metacognitive awareness and take ownership of their learning due to its breadth and clarity.

Winne and Hadwin’s information processing model has four phases: Task definition, Goal setting/planning, Enacting tactics, and Adaptive metacognition \cite{winne2000measuring} and is based on cognitive architecture with a strong analytical detail. This model is suitable for digital learning environments and highly technical for most learners. Boekaerts’ dual-processing introduces emotion and motivation in self-regulated learning and has two pathways: Growth-oriented pathway (learning focused) and Well-being pathway (emotion-focused protection) \cite{boekaerts1999self}. This reflection model is suitable for emotion regulation programs and struggling learners. 

Evidence shows that teacher beliefs shape the promotion of self-regulated learning, as many learners lack robust mental models for effective regulation \cite{lawson2019teachers, bjork2013self}. Self-regulated learning is increasingly supported by digital technologies, including learning analytics, adaptive platforms, and e-portfolios. Self-regulated learning is most dominant in health sciences, STEM education, and professional development contexts. In health sciences, reflective practice is integral to clinical reasoning and professional identity formation, with structured reflection widely implemented in medical and nursing education programs \cite{sinkkonen2024review,phua2024systematic}. In STEM education, self-regulated learning reflection is increasingly embedded in blended and online learning environments to enhance metacognitive awareness and problem-solving skills. Teacher education and professional development also emphasise reflection as a core component for improving instructional strategies and fostering lifelong learning, supported by tools like the self-regulation empowerment program \cite{lawson2019teachers}. 

The integration of GenAI into learning environments introduces unique challenges for self-regulated learning. While self-regulated learning emphasises autonomy, metacognitive monitoring, and iterative reflection, GenAI tools can inadvertently reduce learners’ engagement in these processes by offering instant solutions, which may foster dependency and diminish critical thinking. The complexity of self-regulated learning models relies on active goal-setting and strategy selection, yet GenAI’s automated feedback can bypass these steps, limiting opportunities for self-monitoring and adaptive control \cite{panadero2017review}. Additionally, measuring self-regulated learning becomes more difficult when learners use GenAI-generated content, as traditional self-report instruments fail to capture GenAI-assisted regulation behaviours \cite{greene2007theoretical}. Ethical concerns, such as academic integrity and transparency in reflective practices, further complicate self-regulated learning implementation in GenAI-rich contexts. Therefore, while GenAI offers potential for scaffolding self-regulated learning, its unregulated use risks undermining the very autonomy and metacognitive engagement that self-regulated learning seeks to develop. A summary analysis of selected self-regulated learning models, including their strengths, limitations in the GenAI era, and application areas, is presented in Table~\ref{tab:reflection_models_SRL}.

A review of current reflection practices and models highlights the need for a new integrated framework to incorporate GenAI into learning for the following reasons:
 
\begin{itemize}
    \item Existing models are based on strong theoretical foundation and are suitable for specific educational areas, however they struggle to handle the extensive use of GenAI in education,
    \item A single existing model is unable to consider challenges related to technology penetration, responsible and ethical usage, and emotional factors introduced by GenAI, and 
    \item GenAI challenges the authorship and authenticity of the learner's reflection. Existing reflection models assume that the product is created by the learners themselves. Therefore, there is a need for a reflection model that focuses on the process rather than the product.
    
\end{itemize}
Next, we will review the existing literature, with a primary focus on the need for reflection and the development of new reflection models following the introduction of GenAI tools. 

\section{Reflection Models in the GenAI Era} \label{sec:RW}
This section presents a review of the literature on the reflection published in the last three years, specifically focusing on the critical disruptions and research opportunities catalysed by the introduction of GenAI. 

In \cite{Nemorin2025ActionReflection}, the authors investigate GenAI's potential as a supportive tool for enhancing reflection for moving away from simple, product-focused assignments toward more personalised, process-oriented activities that are less susceptible to superficial, AI-generated responses. Their key contribution is the development of new theoretical and practical frameworks, such as the human-in-the-loop model. These frameworks provide a blueprint for integrating GenAI purposefully, emphasising the irreplaceability of human critical analysis and personal insight. However, there is a lack of longitudinal studies to understand the long-term effects of GenAI integration on metacognitive skill development, along with the lack of widespread empirical validation.

Chan et al.~\cite{Chan2024} argue that the integration of tools like ChatGPT can lead to a transformative shift in education, moving from a teacher-centric to a learner-centric model. The central idea is that AI can be a catalyst for change, forcing a re-evaluation of traditional teaching and assessment methods. Instead of a passive learning experience, AI can facilitate a more active and transformative process. It can act as a personalised tutor, providing instant feedback and support, which can serve as a disorienting dilemma that challenges students' assumptions and prompts critical reflection. The authors explore how AI can be used to redesign curriculum and assessment, fostering higher-order skills like critical thinking, problem-solving, and creativity. The authors emphasise that a key change is embracing AI literacy and creating new policies that guide its ethical and responsible use, ultimately preparing students for a future where human-AI collaboration is essential.

Further, \cite{Chang2025} proposes an AI-assisted method for reflective writing that aims to improve students’ higher-order thinking. Unlike traditional methods where students write unassisted, this approach uses a generative AI tool to provide dynamic prompts and targeted feedback. Modern AI interactions have the capability to guide students from simple recall to complex analysis. The key innovation is an actively supported process that leverages AI as a scaffold, significantly enhancing higher-order thinking skills. The method has three main phases: an initial reflection where students write unassisted; an AI scaffolding phase where the tool analyses the writing and generates tailored questions; and a deeper engagement phase where these prompts challenge assumptions and provoke critical analysis, guiding them to a more complex level of thought.

The scalability and personalisation limits of traditional reflective learning are addressed by proposing the use of Large Language Models (LLMs) as adaptive, accessible facilitators \cite{wang2024generative}. Through an analysis of prompt engineering and simulated multi-turn dialogues, the study demonstrates that pedagogically aligned LLMs can deliver personalised, context-sensitive guidance that effectively promotes deep engagement and critical thinking. However, a significant limitation is the reliance on a simulated, self-play methodology where the LLM acts as both tutor and student, which inherently restricts the external validity and generalisability of the findings to real-world educational settings.

In \cite{Stranges2024Reflective}, the authors examine undergraduate health science students' self-reported use of GenAI on reflective writing, finding that only a minority $(33\%)$ reported using it, largely to enhance learning and efficiency rather than to replace their own reflection. This suggests that reflective writing retains its pedagogical value, as most students used GenAI selectively and ethically, supporting rather than substituting their personal process. However, a significant limitation is the reliance on anonymous, retrospective self-report data, which may be subject to social desirability bias or inaccurate recall, potentially underestimating the true extent of GenAI use or misuse. 

By suggesting a structured self-reflection and declaration form, the article tackles the main issue raised by GenAI in lengthy written examinations \cite{Combrinck2025Student}. The authors found that incorporating this reflective practice supported the ethical use of GenAI and provided lecturers with transparency, noting that students who reflected deeply relied less on AI-generated content. However, the study acknowledged a key weakness: students with high AI-detected scores were often unable to adequately reflect on their usage or provide credible explanations, indicating that self-reflection alone may fail to manage misuse by those who are actively trying to deceive the system. 

Wraith et al. reveal that educators struggled to reliably differentiate between reflections written by medical students and those generated by GenAI, with low to moderate sensitivity and specificity across various reflections \cite{Wraith2024CanEducators}. This highlights a significant challenge to the validity of written reflection as an assessment tool in medical education, suggesting a need to harness GenAI to teach reflective practice skills rather than solely focusing on detection. The major limitation of this study is its small sample size of only 28 educators and a potentially limited set of four reflection types, which may restrict the generalisability of the findings on educator detection capabilities.

In \cite{Fox2025ReflectionAI}, the author explores the dual potential of GenAI in audio production courses, finding that while it can augment creativity, idea generation, and efficiency, educators must grapple with the ethical considerations of authenticity, human agency, and AI literacy. However, as a reflective opinion piece based on the author's own teaching experiences, it lacks empirical student data or a systematic study to quantify the actual impact of GenAI on student creativity or academic integrity. 

In \cite{AlFattal2025YouDoIt}, the author introduces a novel comparative assignment framework that fosters reflective learning by having students contrast their own marketing analyses with GenAI-generated outputs, which successfully encouraged a critical, dialogic engagement with AI rather than passive reliance. However, the study's focus on only a single marketing concept (environmental scanning) and its reliance solely on student self-reflections for data mean that its findings may not generalise to other course topics or accurately capture the true depth of student learning and skill development.

The authors highlight that the individual engagement in critical reflection depends on whether learners perceive conflict or agreement with the output of a machine learning–based clinical decision support system, and the result indicates that the conflict seed can induce deeper reflection than the confirmation seed \cite{abdel2023ai}. 

The key observations from the literature review are: 
\begin{itemize}
     \item There is greater demand for the use of reflective learning after the induction of GenAI to ensure the learning, and 
    \item  Existing reflective models do not have clear strategies, and no new reflection model is introduced to tackle the problem introduced by GenAI. 
 \end{itemize}
Therefore, there is a need for a reflection model that addresses the increased demand in reflection in the era of GenAI, that addresses the limitations of the existing reflection models, and also considers the technological aspects of GenAI.

\section{Model Primer}\label{sec:primer}

As discussed in the previous sections, the shift toward GenAI-integrated learning environments has created a pedagogical gap that traditional reflection models cannot bridge. While these existing models are theoretically robust for human-only contexts, they lack the structural transparency required to assess agency and authenticity in the emerging era of GenAI. 
However, it should be noted that the influence of GenAI is not inherently detrimental; studies indicate that, within a structured setting, GenAI can significantly enhance metacognition, higher-order thinking, and guided contemplation of the learner \cite{elsayary2024integrating}.

To address the complexities and existing challenges, there is a need for a new reflection model that offers a methodical, technologically aware approach to reflection that maintains human components of judgement, ethical reasoning, and critical thinking, even when students employ GenAI tools for learning. The reflective model needs to view GenAI as a scaffold that supports students' cognitive processes rather than as a threat, making sure that their contributions, choices, and reflective reasoning are transparent, observable, and assessable. 

There is a need for a new reflection model, which can provide learners with an opportunity to produce true reflection while GenAI is used for learning. To ensure this new model is both theoretically robust and practically relevant, a systematic two-stage development process has been adopted. The first stage involves a critical diagnosis of why current reflection frameworks are insufficient for the GenAI era. The second stage moves into a constructive phase, synthesising established learning theories into a new integrated framework. This methodological progression ensures that the proposed model is not merely a reaction to new technology, but a robust evolution of existing pedagogical practices.
\subsection{Problem Identification and Theoretical Grounding}
To establish a rigorous foundation for the new model, this paper uses \textit{Design-Based Research (DBR) methodology} \cite{dbrc2003design} assisted by \textit{Critical Inquiry approach} \cite{brookfield1995becoming} to study the requirements, essential components analysis, and development. This process helps us critically identify the problem and the requirements for the new reflection model. The problem is analysed critically by the following process across the existing literature.
\begin{itemize}
    \item Purpose: To critically examine existing reflection models and their theoretical perspectives to identify how they fail to account for the influence of GenAI. 
    \item DBR Function: Problem identification and theoretical grounding.
    \item Critical Inquiry Lens: Question assumptions about authenticity, agency, and authorship in reflective learning when GenAI is present.
    \item Research Questions:
        \begin{itemize}
            \item In what ways do existing reflection models address or overlook the influence of GenAI on reflective thinking and representation?
            \item What are the gaps emerging when traditional notions of reflection are examined in the presence of GenAI?
        \end{itemize}
\end{itemize}

The critical analysis was conducted based on the above research questions, along with the detailed literature survey in Section~\ref{sec:TF} and Section~\ref{sec:RW}. The literature analysis, in Section~\ref{sec:TF}, confirmed that traditional reflection models, while theoretically robust in human-only contexts, lack the necessary structure and sensitivity to manage GenAI's influence on artefact production, authenticity, and ethical practice. The core inadequacy lies in their failure to make the GenAI-aware cognitive process transparent and to integrate an explicit focus on technological awareness and ethical reasoning. 

In Section~\ref{sec:RW}, we conducted a review of the recent literature for the need for reflection and any recent development of the reflection model in the era of GenAI. This review establishes that there is a greater demand for reflection in learning while using GenAI, but no new model has been developed that adequately addresses the issues discussed before.     

The findings of the analysis indicate that there is a need for a new reflection model that utilises the theoretical soundness of the existing models and also addresses GenAI technological awareness. Consequently, the second stage of the model development process focuses on identifying the attributes and features that need to be adopted into a cohesive framework to address technological awareness and meet industry demands.

\subsection{Model Design and Development}
To create a reflection model that is both adaptable to GenAI and prioritises human judgement, we argue that there is a need for an iterative and integrated reflection model that strategically fuses five essential theoretical components derived from established learning and reflection theories. After a detailed analysis of  existing  learning and reflection theories (refer to Section~\ref{sec:TF}), the following key features are adopted from existing reflection and learning practices in the literature:

\begin{figure}[htp]
\centering
\includegraphics[scale=0.450]{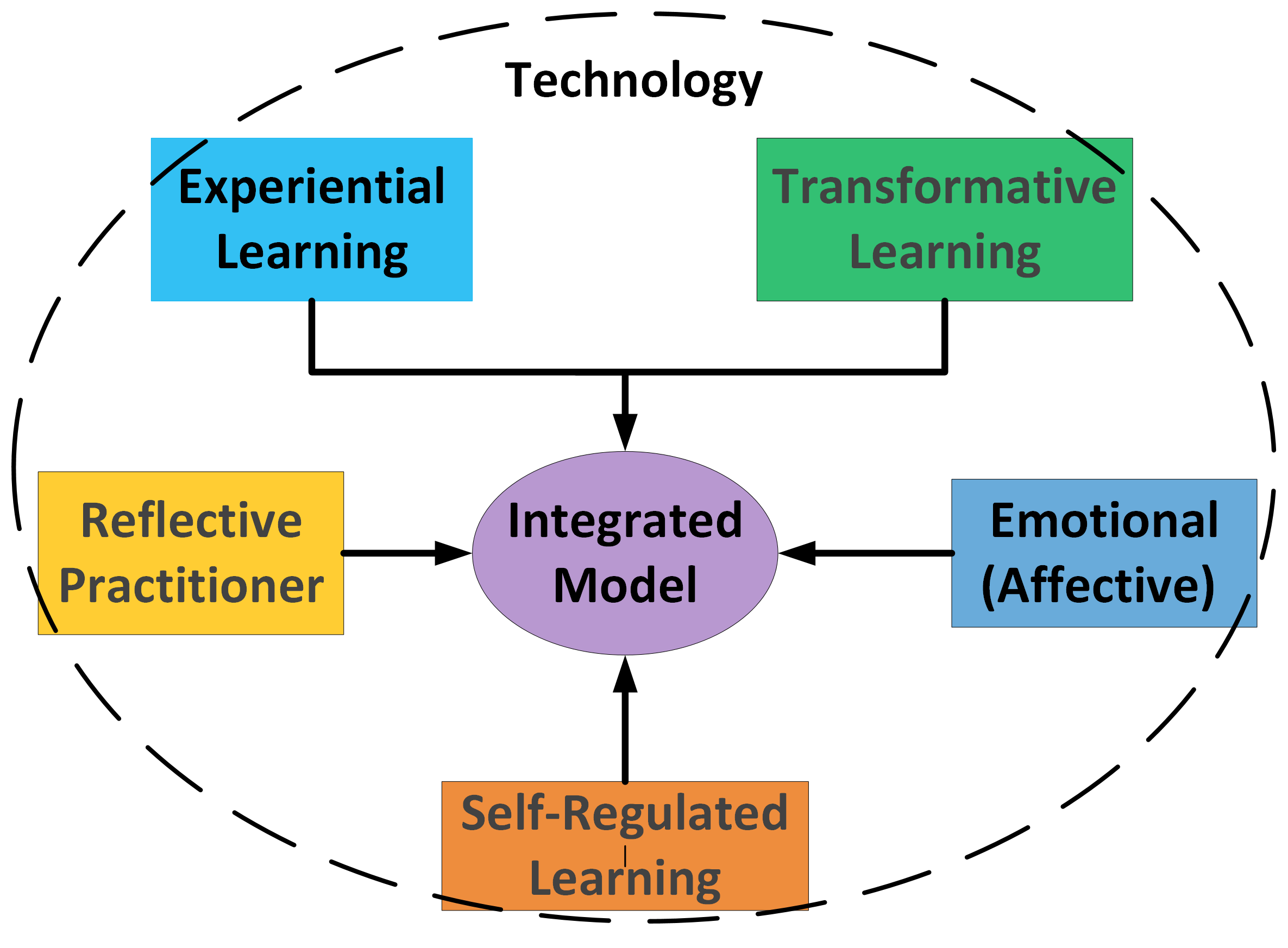}
    \caption{Integrated self-reflection framework with core theoretical foundations and a technology-enhanced learning layer.}
     \label{fig:NRmodel}
\end{figure}

\begin{itemize}
    \item \textbf{Experiential Learning:} This reflection model focuses on concrete experience. Engaging with GenAI tools (e.g., prompting with GenAI) is treated as an active, concrete experience to be reflected upon. Structured reflection stages and an iterative learning cycle are key features adopted from these experiential based reflection models.
    
    \item \textbf{Transformative Learning:} Emphasises deep reflection on assumptions and disorienting dilemmas (like an AI output challenging a belief), critical for perspective transformation. Therefore, there is a need for transformative learning to re-examine their own assumptions, beliefs, and habitual ways of
    understanding the world.

    \item \textbf{Reflective Practitioner:} From these models, continuous reflection during and after the task, particularly Reflection-in-Action (e.g., iterative prompting and refinement) and Reflection-on-Action (reviewing and analysing the final output), are key features adopted from these reflective practitioner based models.

    \item \textbf{Emotional (Affective):} While using GenAI in education, learners require acknowledging emotions, values/beliefs, and well-being, vital for navigating the ethical and personal implications of delegating cognitive tasks to GenAI. Therefore, the new model should systematically handle emotion factors. 
    
    \item \textbf{Self-Regulated Learning:} These models provide the necessary structure for students to manage their learning with GenAI, focusing on goal setting, self-monitoring, strategic planning, adaptive adjustment, and autonomy. These features are adopted in the new model.
\end{itemize}

The integrated reflection framework is illustrated in Fig.~\ref{fig:NRmodel}, which provides the conceptual structure for integrating input from various reflective theories and practices, as well as GenAI technological aspects. The next section presents an implementation of an integrated framework, operationalised through the 5P model, which identifies how each stage of the model addresses the theoretical aspects highlighted in the integrated framework.

\section{5P Reflection Model} \label{sec:P5}
\begin{figure}[htp]
    \centering
    \includegraphics[scale=0.28]{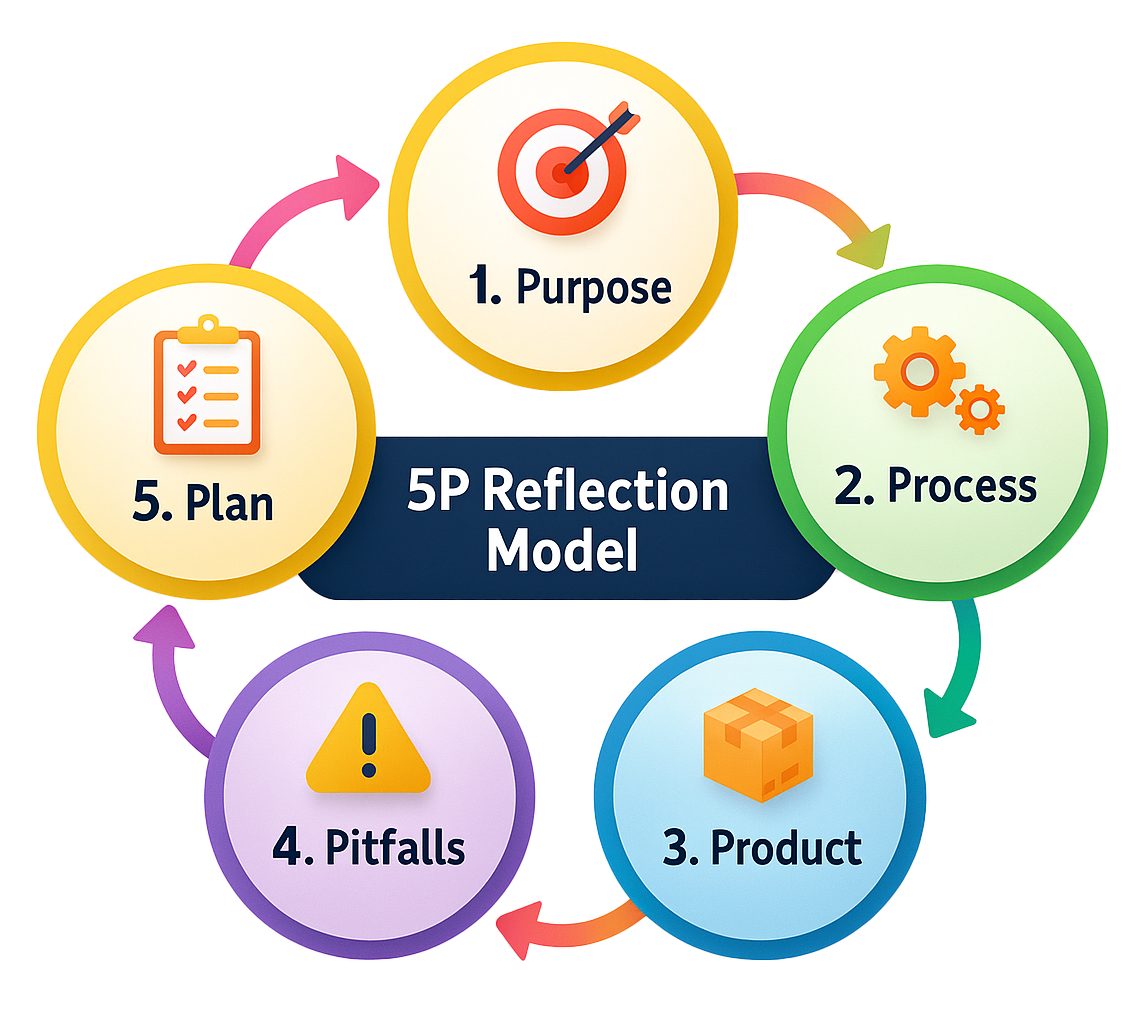}
    \caption{5P reflection model for the GenAI era.}
     \label{fig:p5model}
\end{figure}

The 5P Reflection Model (illustrated in Fig.~\ref{fig:p5model}) is an iterative reflective model with five stages: \emph{P}urpose, \emph{P}rocess, \emph{P}roduct, \emph{P}itfalls, and \emph{P}lan. The model has adopted various features from different reflective theories and practices at different stages, with emotional factors considered across all stages. The model also incorporates the technological dimensions of GenAI and its responsible use in learning environments.
\subsection{Purpose}
This initial stage of the reflection model is fundamentally driven by principles of self-regulated learning, focusing on the \textit{forethought} \cite{zimmerman1990self}. Learners take ownership of their learning and the use of GenAI tools by setting clear goals and objectives that meet academic and institutional standards. Before any learning or GenAI usage commences, the learner must establish autonomy and responsibility by taking full ownership of their educational journey and tool choices. This involves rigorous goal-setting, in which learners define clear, achievable objectives for both their learning outcomes and their use of GenAI tools. The questions discussed here are essential for laying this foundation by clarifying the learning objectives and the intended use of the GenAI tool in advance. The learners may articulate that the purpose of the usage of GenAI: enhance efficiency, boost creativity, achieve clarification, or any other aims. This navigates learners to use GenAI in a deliberate and planned approach ensuring that the learning process is focused and intentional rather than reactive or aimless. This is crucial, as the use of GenAI in education shifts learning toward a more active learning environment \cite{Chan2024}.

The learners need to make sure that the usage of the GenAI tools follows the assessment requirements, guidelines, and institutional policies. This choice should be informed by a precise clarification of the intention of the use (e.g., idea generation, drafting, coding support, research aid, information gathering and so on). By setting these goals upfront, the learner establishes a clear standard for later evaluation, ensuring the GenAI usage is a strategic aid for learning rather than a replacement for learning. Furthermore, this forethought integrates necessary checks to ensure the work will meet academic and institutional standards, and that the learners will remain ethical and responsible users of the technology. This preparatory phase guarantees that GenAI is employed thoughtfully and responsibly, aligning with the overall educational objectives from the very beginning \cite{unesco2023guidanceAI}.

The learner's initial emotions (excitement, anxiety, fear of failure, or boredom) directly impact their goal setting and commitment. If learners feel anxious about a complex task, they might over-reliance on GenAI from the start, setting an inadequate learning goal and usage of GenAI tools. Feelings of empowerment or helplessness influence the learner's ability to take ownership. A learner who feels overwhelmed may delegate the entire intellectual task to the GenAI, undermining the foundation of the learning. Therefore, balanced emotional support from educators is essential for sustaining learners’ motivation and self-efficacy while using GenAI.

\subsection{Process}
This phase centres on the active deployment of the GenAI tool as an assistive tool to achieve learning goals established in the previous phase. This phase is heavily influenced by the principles of the Reflective Practitioner \cite{schon1983reflective,Schon1987}, emphasising \textit{Reflection-in-Action}. The critical thought and process that occurs while prompting with GenAI, where learners must detail how effectively they use GenAI tools and document the detailed steps taken (e.g., prompting, iterations, and evaluation of outputs). The learners' capacity to reconsider, modify, and adjust their choices of prompts and outputs as circumstances evolve is key to handling GenAI tools effectively. Additionally, the learner acknowledges and understands that the outcomes of GenAI are probabilistic in nature and entirely dependent on the input (prompt), as well as the GenAI model itself. Therefore, a key focus is on how learners refine or iterate their interaction with the tool. The learner needs to record the specific prompts or instructions given and document any adjustments made while using GenAI. This iterative process is the core of effective GenAI usage, requiring continuous monitoring and adaptation.

The necessity of validation of the GenAI's output is paramount due to its probabilistic nature. This demands focused reflection-in-action while both formulating prompts and assessing the generated content. Learners must specifically address the verification of GenAI’s output to maintain academic integrity. Emotions directly modulate cognitive function, and frustration or impatience during the iterative prompting process can lead the learner to stop refining their input prematurely or accept a low-quality output, compromising the entire process. Feelings of relief or satisfaction upon receiving a seemingly complete GenAI output can lead to confirmation bias and insufficient cross-checking. Monitoring these emotional responses reminds the learner to maintain necessary critical human judgement. Educator's encouragement to learners to monitor emotional responses during learning and reflect on the interplay of cognitive, affective, and contextual elements is critical. 

\subsection{Product}
In this stage, the learner engages in rigorous \textit{Reflection-on-Action}, a principle from the Reflective Practitioner model \cite{schon1983reflective,Schon1987}, by critically reviewing and analysing the outcome and process after the learning activity is completed using GenAI. The primary focus is on the outcome and its quality. The learner must evaluate the quality, originality, and accuracy of the final product, paying close attention to the core instruction: \textit{Always validate the product from GenAI with other reliable sources}. This cross-validation is essential for ensuring academic rigour and confirming that the potentially probabilistic output from the GenAI tool is correct and reliable. This process ensures that learners have full confidence that the final product meets necessary standards and trustworthiness.

Learners must be able to demonstrate human contribution in the product generation to ensure full intellectual ownership. This is achieved by directly addressing the question of what contributions come from learners and what contributions are generated by GenAI. Distinguishing these roles emphasises the learners’ input and their judgement. By analysing the contributions against the initial learning goals (set in the purpose stage), the learner assesses the true impact of the assistive tool. A key reflective consideration here is: Did GenAI enhance or limit learners' creativity in the product development? This critical self-assessment goes beyond mere task completion, focusing on the deeper educational value of the GenAI interaction. It ensures the tool was used ethically and effectively to support rather than replace the learner’s creative input and learning process.

Learners' emotions tied to the outcome or product quality (e.g., pride, disappointment) can bias the learner's acknowledgement of human and GenAI contribution. A desire for self-recognition might lead a learner to downplay the role of GenAI. At the same time unfamiliarity with the tool might lead to unnecessary criticism on its capabilities and usefulness. The affective state after completing the task dictates the honesty of the reflection. If the learner feels stressed or rushed, their evaluation of quality, originality, and accuracy may be superficial. Therefore, educators' supportive and non-judgemental guidance is crucial to help learners navigate these emotional biases and ensure the authenticity and depth of their reflection.

\subsection{Pitfalls}
This phase deals with the inherent pitfalls of GenAI technologies, which are a crucial component for responsible and ethical GenAI usage in the learning cycle. Learners must evaluate the challenges or risks faced while using GenAI in various aspects: Factual and cognitive; Ethical and integrity; and Privacy and security. Factual and cognitive risks include hallucinations (GenAI generating false information with confidence); Ethical and integrity risks encompass plagiarism, bias, and the temptation towards over-reliance on GenAI; Privacy and security risks associated with data input to GenAI. It is vital for learners to actively reflect on the moments where critical human judgement was essential to correct the GenAI's output or to interpret ambiguous results. Furthermore, the learner must distinguish between appropriate and inappropriate use to ensure the tool is genuinely assistive and not hindering deep learning. This constant internal scrutiny solidifies the learner's role as the final, responsible authority over the generated content. By reflecting on these factors, the learner develops a nuanced, ethical, and informed perspective on GenAI. This process ensures that the learner's interaction with the tool is not just technically sound but is also conducted with full ethical awareness.

The temptation to over-reliance on GenAI is an emotional one, often driven by the desire for efficiency, ease, or avoiding hard work. Managing and reflecting on this emotional drive is necessary for developing genuine ethical self-regulation. Recognising ethical concerns (like plagiarism or bias) often starts with a feeling of unease or conflict. Therefore, there is a need to address ethical discomfort, and monitoring this affective response is the first step toward engaging ethical reasoning and making appropriate choices while using GenAI in learning. 
 
\subsection{Plan}
In this stage, learners are required to synthesise insights from all preceding stages to inform future actions, driving genuine transformative learning. The first step involves a critical analysis of progress based on the objectives set in the Purpose Stage, answering the key question: Did I achieve my initial learning and GenAI usage goals? This reflective analysis facilitates the evaluation of personal growth, allowing the learner to identify where skills were strengthened and where reliance on the tool was appropriate or excessive. The activities during the Process and Product Stages provide an evaluation of the efficiency in the GenAI usage and learners' contributions. By reflecting on the Pitfall Stage, the learner gains deep insights about using GenAI responsibly, which forms the foundation for developing a sophisticated, self-aware approach to future digital tool integration.

The ultimate objective of this stage is to create a clear pathway for continuous improvement in the use of GenAI in learning. The learners must articulate how they will adjust their approach in the future by setting new, refined goals for GenAI interaction. This plan outlines practical strategies to help learners use GenAI in ways that support learning and personal skill development, ensuring that these tools enhance rather than undermine core competencies. This prospective action plan ensures that the learning experience is not an isolated event but contributes to transformative learning, permanently altering the learner's understanding and application of digital tools in a responsible, ethical, and academically sound manner. The sustained feeling of confidence and insight gained through the reflection process is essential for the learner to successfully adjust their approach in the future and maintain the balance between GenAI support and personal skill development.

These five (5) stages of the proposed 5P reflection model together provide a comprehensive solution to the issues raised through the research questions. The 5P Reflection Model offers a robust, integrated approach that successfully bridges the gap between traditional reflective practice and the realities of GenAI-assisted learning. By aligning each stage, from the initial intentionality of \textit{Purpose} to the forward-looking \textit{Plan}, the model ensures that GenAI provides a structured scaffold rather than a substitute for student thought. By integrating the \textit{Pitfalls} stage as a core component and embedding emotional monitoring throughout the cycle, the model moves beyond traditional frameworks that fail to account for the probabilistic and persuasive nature of GenAI. This five-stage framework embodies the philosophy of ``process over product'' by providing a clear, actionable pathway for including GenAI in the curriculum, allowing educators to foster deep, authentic reflection that is both observable and assessable. To ensure this model reaches its full potential in practice, the following section discusses the specific considerations for its adoption and the necessary scaffolding for effective implementation.

\section{Adoption Considerations} \label{sec:Discussion}

The model has been developed with a strong theoretical foundation that integrates experiential learning, self-regulated learning, reflective practice, transformative learning and the emotional aspect of the learner. The features of self-regulated learning used in the Purpose stage ensure metacognitive planning, while the features from the reflective practitioner used in the Process and Product stages promote critical reflection during and after the action with GenAI. In pre-action planning (in the Purpose stage), setting goals and objectives mandates intentionality of GenAI usage by making the learner define the ``why'' of using GenAI, ensuring it is used as a strategic aid. 

The model directly addresses GenAI risks like bias, plagiarism, ethical issues, and over-reliance and prevents over-reliance on technology while aligning specific learning goals. In-action monitoring (in the Process stage) addresses the technical nature of GenAI by focusing on prompting, iteration, and validation. Post-action evaluation (in the Product stage) requires rigorous validation of the final output against external sources, acknowledging that the probabilistic nature of GenAI outputs is crucial for the responsible use of GenAI. A dedicated Pitfalls stage explicitly handles ethical and risk factors which is a key strength of the proposed model. This ensures that the learner doesn't just evaluate the quality of the output but the consequences of the interaction with GenAI. The model emphasises the critical human judgement in various stages (e.g., prompt refinement, verifying output, distinguishing appropriate use, resisting over-reliance) during learning, which checks against GenAI limitations. 

Certain aspects need to be considered while adopting this model for reflection. The model provides a comprehensive and structured framework that facilitates deep introspection at every critical stage of the learning process i.e., before, during, and after the task. This exhaustive approach is specifically designed to prevent the 'surface-level' reflection often found in GenAI interactions. By requiring detailed documentation and consistent check-ins, the model ensures that reflection remains a substantive cognitive exercise, transforming even routine tasks into opportunities for higher-order critical thinking. Therefore, the model is generally suitable for a major assessment, capstone project or professional development evaluation. Secondly, the effective implementation of this model requires explicit instruction and scaffolding from educators. Learners must be taught how to monitor emotional responses, design and refine prompts iteratively, and cross-validate the output. Educators must possess the necessary skills to deliver clear and effective scaffolding which can be a significant challenge and highlights the need for ongoing professional development. 

Like other reflection models, this model also has limitations in assessing emotional aspects of learning. Assessing the quality of internal thoughts, such as monitoring emotional responses or reflecting on the temptation to over-reliance (in Process and Pitfalls stages), is highly subjective and difficult to evaluate. The model relies heavily on the learner's honesty and metacognitive capacity in this aspect. 
Lastly, the Process stage requires documenting specific prompts and iterations, which can be challenging to capture and document accurately. Specialised tools need to be developed to capture learners' continuous interaction with GenAI. 

\section{Conclusion}\label{sec:conclusions}
GenAI is rapidly emerging across multiple fields, including education. As learning and professional development increasingly occur in GenAI-aware environments, reflective practice remains a critical approach for evaluating learners’ cognitive and professional growth. However, traditional reflective models are increasingly inadequate in addressing the range of challenges introduced by GenAI-enabled learning environments. In response, this paper introduces a new reflective model, called the 5P (Purpose, Process, Product, Pitfall, and Plan) model, developed using an integrated conceptual framework. The proposed model also incorporates explicit consideration of the strengths and limitations of GenAI technologies, strengthening learners’ critical and metacognitive reflection. In addition, the paper outlines practical considerations for adopting and implementing the proposed model in educational and professional development contexts.

\bibliographystyle{IEEEtran}
\bibliography{Myref.bib}
\section*{Biography Section}
\begin{IEEEbiography}[{\includegraphics[width=1in,height=1.25in, clip,keepaspectratio]{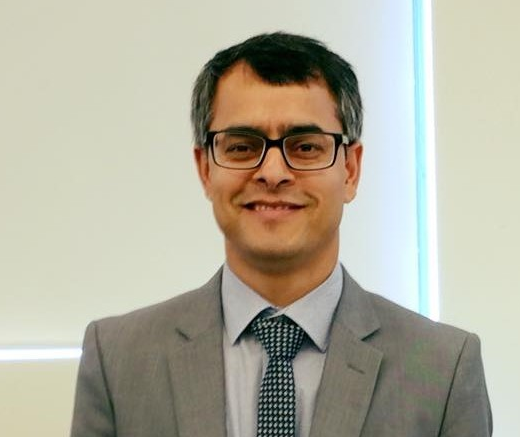}}]{Rajan Kadel} received his B.Eng. degree in Computer Engineering from Tribhuvan University, Nepal, in 2002, followed by an M.Sc. degree in Telecommunications Engineering from the University of Gävle, Sweden, in 2007. He earned his PhD in Telecommunications Engineering from Adelaide University, Australia, in 2013. He is currently an Associate Professor and Head of School at the National Academy of Professional Studies (NAPS), Australia. Dr Kadel brings over two decades of experience in teaching, research, and professional practice. Before his academic career, he worked extensively in the telecommunications industry, serving as a Switching Supervisor at Nepal Telecom and as an Assistant Manager at the Nepal Telecommunications Authority. His research interests include teaching methodologies, error-control coding, Wireless Sensor Networks, and Wireless Body Area Networks.
\end{IEEEbiography}

\vspace{-33pt}
\begin{IEEEbiography}[{\includegraphics[width=1in,height=1.25in, clip,keepaspectratio]{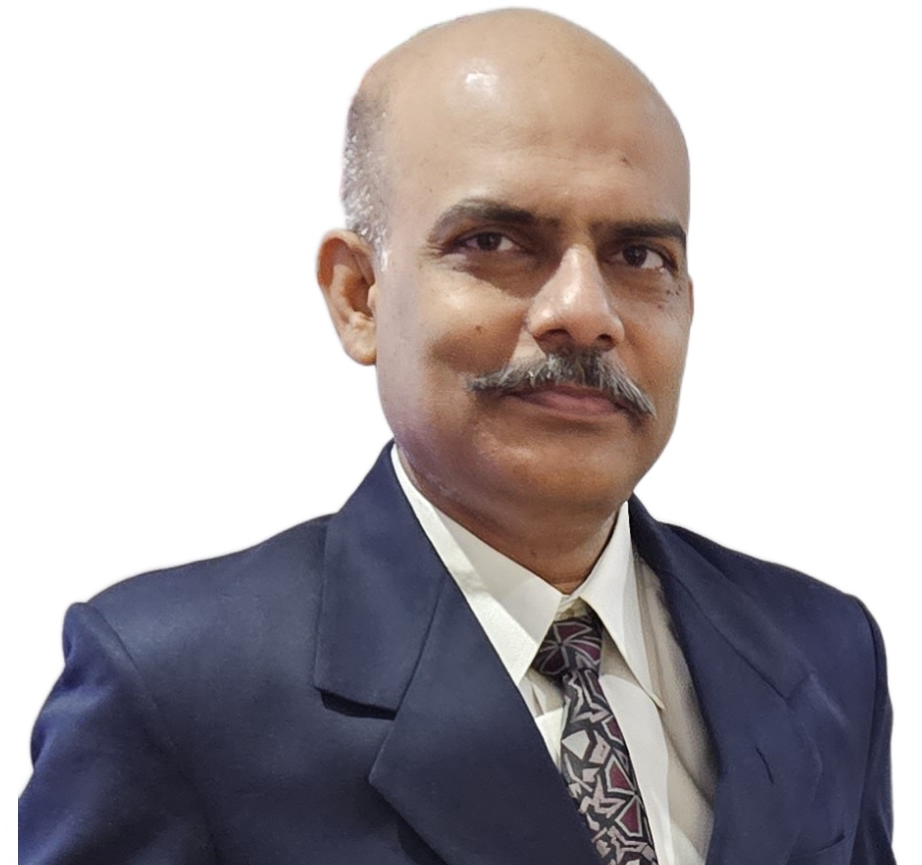}}]{Samar Shailendra} (Senior Member, IEEE) received his PhD from IIT Guwahati. He is currently working as a Senior Lecturer at the Melbourne Institute of Technology, Australia. He is also a Visiting Professor at IIIT Bangalore. Previously, he worked with Intel and led their Mobile Edge Computing (MEC) Standardisation at 3GPP and India. He also served as the chair of the TSDSI Roadmap Committee and vice chair of Study Group - Networks (SGN) at TSDSI. He is credited with making substantial contributions to shaping 5G Advanced requirements and the global vision for 6G technology in close collaboration with Indian Telecom Standard Bodies and Government Agencies. His research interests include SDN/NFV, Internet Architecture, Transport Protocols, M2M, Drones, Robotics, AI and Quantum Computing.
\end{IEEEbiography}
\vspace{-33pt}
\begin{IEEEbiography}[{\includegraphics[width=1in,height=1.25in, clip,keepaspectratio]{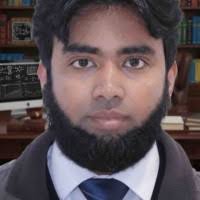}}]{Mohammad Tahidul Islam} received his PhD from RMIT University, Australia, in 2021. He is currently working as a lecturer and unit coordinator at Melbourne Institute of Technology (MIT), Melbourne, Australia. He was actively engaged in research, supervision and teaching for the last twelve years at RMIT University (Australia), Ulsan University (South Korea) and International Islamic University Chittagong (Bangladesh). He dedicated his expertise to contribute on the next generation wireless communication, such as Cognitive radio communication, NextG Communication and Networking, 5G Satellite Communication, Smart Grid Communication System, Cloud Engineering for Internet of Things (IOT), Statistical data analysis and 5G wireless communication, Artificial Intelligence and Machine Learning.
\end{IEEEbiography}
\vspace{-33pt}
\begin{IEEEbiography}[{\includegraphics[width=1in,height=1.25in, clip,keepaspectratio, angle=270]{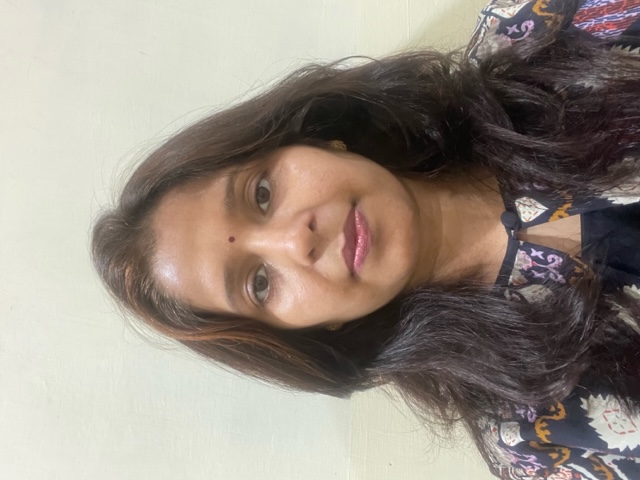}}]{Urvashi Rahul Saxena} is a seasoned Academic Professional with over 20 years of teaching experience. She is currently working in the School of IT and Engineering (SITE), Melbourne Institute of Technology (MIT), 288 La Trobe Street, Melbourne, VIC 3000, Australia. She completed her PhD in CSE from Jaypee Institute of Information Technology, Noida, India and her M. Tech in Computer Science from Birla Institute of Technology, Mesra, Ranchi, India and her B.E in Information Technology from Rajiv Gandhi Proudyogiki Vishwavidyalaya, Bhopal. Her research areas include Cloud Computing, Restricted Access Control, Trust Computation, Distributed Computing, and Networking. 
\end{IEEEbiography}
\vspace{-33pt}
\begin{IEEEbiography}[{\includegraphics[width=1in,height=1.25in, clip,keepaspectratio]{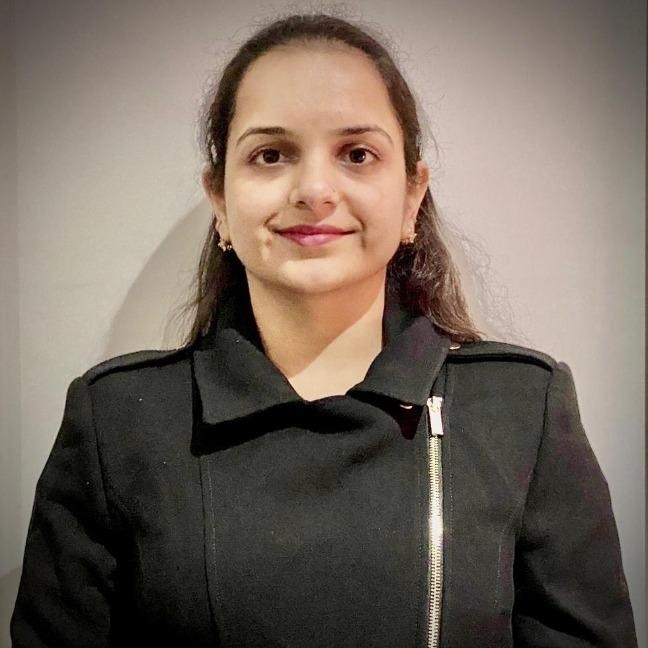}}]{Aakanksha Sharma} received her PhD from Federation University, Australia, in 2022.  She currently serves as a Lecturer at Melbourne Institute of Technology. She has a strong teaching background and has worked at Federation University and Royal Melbourne Institute of Technology (RMIT). Before this, she worked as an Assistant Professor at Chandigarh University for five years. Her research interests include the Internet of Things, Software-Defined Networks, wireless communications, Artificial Intelligence, and Quantum Computing.
\end{IEEEbiography}
\vspace{-33pt}
\begin{IEEEbiography}[{\includegraphics[width=1in,height=1.25in, clip,keepaspectratio]{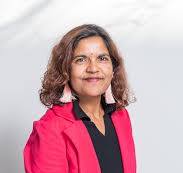}}]{Sabitra Kaphle} has over 20 years of international experience in research, education, gender equity, and health policy. She is a qualitative and cross-cultural researcher focusing on culturally safe and effective interventions for low-resource, socially complex, remote, regional, and disadvantaged communities. Her research work centres around amplifying the voices of vulnerable groups to promote equity and access to education, economic resources, and health services. 
Dr Kaphle's research further explores the intersections of gender, social inequities, health, culture, and structural disadvantage. She also serves in advisory roles on gender-based policy and health systems across multiple countries. Dr Kaphle is widely published in her areas of expertise, including her book Socio-cultural Insights of Childbirth in South Asia (Routledge, 2022). She currently works as a Senior Lecturer with the School of Health, Medical and Applied Sciences at Central Queensland University, Australia.
\end{IEEEbiography}

\end{document}